\documentclass[%
 aip,
rsi,%
 amsmath,amssymb,
 reprint,%
]{revtex4-1}

\usepackage[dvipsnames,table]{xcolor}
\usepackage{xcolor}
\usepackage{ulem}
\usepackage{graphicx}
\usepackage{dcolumn}
\usepackage{bm}
\usepackage{nicefrac}
\newcommand{\n}[1]{\mathbf{#1}}

\begin{document}

\preprint{AIP/123-QED}

\title[Rodr\'{i}guez et al.]{Generalized Boozer coordinates: a natural coordinate system for quasisymmetry}

\author{E. Rodr\'{i}guez}
 \altaffiliation[Author to whom correspondence should be addressed: ]{eduardor@princeton.edu}
 \affiliation{ 
Department of Astrophysical Sciences, Princeton University, Princeton, NJ, 08543
}
\affiliation{%
Princeton Plasma Physics Laboratory, Princeton, NJ, 08540
}%

\author{W. Sengupta}
 \affiliation{ 
Department of Astrophysical Sciences, Princeton University, Princeton, NJ, 08543
}
\affiliation{%
Princeton Plasma Physics Laboratory, Princeton, NJ, 08540
}%

\author{A. Bhattacharjee}
 \affiliation{ 
Department of Astrophysical Sciences, Princeton University, Princeton, NJ, 08543
}
\affiliation{%
Princeton Plasma Physics Laboratory, Princeton, NJ, 08540
}%

\date{\today}

\begin{abstract}
We prove the existence of a straight-field-line coordinate system we call \textit{generalized Boozer coordinates}. This coordinate system exists {{} for magnetic fields with nested toroidal flux surfaces} provided $ \oint\mathrm{d}l/B\:(\mathbf{j}\cdot\nabla\psi)=0$, where symbols have their usual meaning, and the integral is taken along closed magnetic field lines. {{}All quasisymmetric fields, regardless of their associated form of equilibria, must satisfy this condition}. This coordinate system presents itself as a convenient form in which to describe {{} general} quasisymmetric configurations and their properties. Insight can be gained analytically into the difference between strong and weak forms of quasisymmetry, as well as axisymmetry, and the interaction of quasisymmetry with different forms of {{} equilibria}.
\end{abstract}

\maketitle

\section{Introduction:}\label{sec:intro}

Quasisymmetric stellarators\cite{nuhren1988,boozer1981,rodriguez2020} (sometimes referred to as helically-symmetric\cite{tessarotto1995,tessarotto1996}) constitute an attractive choice for magnetic confinement fusion. Theoretically, such designs exhibit transport properties analogous to those of axisymmetric devices\cite{boozer1983} while possessing larger three-dimensional freedom. The latter allows avoiding some of the inherent limitations of tokamaks. The quasisymmetric stellarator achieves this by possessing a hidden symmetry, namely, the magnitude of the magnetic field $|\mathbf{B}|$ is symmetric, while the full vector $\mathbf{B}$ is not. \par
The concept of quasisymmetry (QS) is elegant, but it appears to have a significant theoretical limitation. It was soon realized\cite{garrenboozer1991b} that constructing a configuration with exact QS is not possible. Although there is no definitive proof that shows this impossibility, work on near-axis expansions\cite{garrenboozer1991b,landreman2018a} supports this point of view. The governing system of equations is overdetermined: there are more constraints than degrees of freedom. This limitation does not, however, prevent designs that exhibit behavior close to QS in a volume.\cite{bader2019,henneberg2019} \par
Recently, studies that explore the concept of QS more deeply have appeared.\cite{rodriguez2020,burby2021} The main idea has been to separate the concept of QS from that of macroscopic force balance. In these studies, QS is defined as a property of the configuration that confers the single-particle dynamics an approximately conserved quantity without making any statement about the form of macroscopic equilibrium. This perspective differs significantly from the standard approach. Prior to [\onlinecite{rodriguez2020}] and [\onlinecite{burby2019}] {{} (and with very few exceptions\cite{tessarotto1996,burby2013})}, the concept of QS was framed in the context of magnetohydrostatic (MHS) equilibria satisfying $\mathbf{j}\times\mathbf{B}=\nabla p$, where $\mathbf{j} {{}=\nabla\times\mathbf{B}}$ is the current density and $p$ is the scalar pressure. As MHS is not intrinsic to QS, it is important to define QS without reference to a particular form of equilibrium. Separating QS from equilibrium allows us to investigate more deeply its meaning and limitations.   \par
Abandoning the convenient form of MHS equilibrium, although conceptually appropriate, comes at a cost. The magnetic field no longer needs to satisfy the condition $\mathbf{j}\cdot\nabla p=0${ {}, implicitly assumed in most of the literature\cite{nuhren1988,boozer1983,tessarotto1995}}. Hence, one cannot guarantee the existence of Boozer coordinates\cite{boozer1981} as presently understood, even if magnetic flux surfaces exist. Boozer coordinates are particularly convenient for studying QS, as it presents the symmetry in an explicit, simple form. This leads to a search for an analogous, convenient, but more general coordinate system for quasisymmetric configurations.  \par
In this paper, we construct such a coordinate system, which we call \textit{generalized Boozer coordinates} (GBC). This coordinate system was used to formulate the near-axis expansion in [\onlinecite{rodriguez2020i}], in which a QS equilibrium with anisotropic pressure was shown to avoid the conventional problem of overdetermination. The present paper is organized as follows. First, we introduce, develop and discuss this coordinate system systematically and rigorously. We start by presenting a constructive proof for GBC and the class of fields for which it exists. {{} We then present the set of equations describing a quasisymmetric magnetic field in this coordinate system.} This gives us the opportunity to gain an alternative\cite{burby2019,rodriguez2020} perspective on the distinction between the so-called weak and strong forms of quasisymmetry, as well as a comparison to axisymmetry. We close by summarizing the equations that link the equilibrium and the quasisymmetric field and concluding remarks.

\section{Generalized Boozer coordinates}

\subsection{Explicitly symmetric formulation of quasisymmetry} \label{sec:GBCQS}
Let us start this section by introducing the notion of QS. We consider QS from the recent general perspective of single particle motion.\cite{rodriguez2020,burby2021} QS (and in particular \textit{weak} QS) is the property of the fields that confers the motion of guiding-centre single particles with an approximately conserved quantity.\cite{rodriguez2020} For the dynamics of particles to exhibit this conservation it is necessary for the magnetic field to have nested flux surfaces $\mathbf{B}\cdot\nabla\psi=0$ and satisfy,
\begin{equation}
    \nabla\psi\times\nabla B\cdot\nabla\left(\mathbf{B}\cdot\nabla B\right)=0. \label{eqn:tripleQS}
\end{equation}
This form of quasisymmetry is most commonly referred to as the \textit{triple vector product formulation}. Although this is not the form that comes directly from the single-particle analysis (that form is the one used in Sec.~IIIA), it is a succinct way to impose QS on a magnetic field. \par
{{} Given that this generalized concept of QS has a single particle origin, no notion of macroscopic equilibrium is involved. Of course, from a practical point of view, any steady field of interest will be in some form of force balance. With the definition of QS in (\ref{eqn:tripleQS}), different forms of equilibria may be investigated to understand how they interact with QS. One may generally refer to the macroscopic forces by $\mathbf{F}$, and we do not attempt to assess their origin in this paper. Instead, we focus on the requirements at a fluid level. A complete view of the problem would require an investigation of the kinetic basis of the forces to link the fluid forces to microscopic behavior. This is particularly important as microscopic forces could break the symmetry (and with it, the QS-related conserved momenta). An important example is the electrostatic potential, whose symmetry is needed to the appropriate order and imposes constraints even on the forces that arise from the plasma, such as centrifugal force or anisotropic pressure. With this in mind, we shall focus on the macroscopic aspects.}   \par
Looking at the statement of QS in the form presented in Eq.~(\ref{eqn:tripleQS}), the existence of a symmetry in the problem is not apparent. In the usual context of $\mathbf{j}\times\mathbf{B}=\nabla p$, this absence of obvious symmetry can be amended by using Boozer coordinates. In these coordinates, a field in magnetostatic equilibrium with well-defined flux surfaces is quasisymmetric (aside from quasipoloidal symmetry) iff,
\begin{equation}
    B=B(\psi,\theta-\Tilde{\alpha}\phi), \label{eqn:BQSBooz}
\end{equation}
where $\{\psi,\theta,\phi\}$ are Boozer coordinates, $\psi$ the flux surface label, $\theta$ and $\phi$ poloidal and toroidal angles respectively, and $\Tilde{\alpha}=N/M|M,N\in\mathbb{Z}$. In order to employ Boozer coordinates, the $\mathbf{j}\cdot\nabla\psi=0$ property of MHS equilibria is central. Boozer coordinates {{} have the standard straight field-line coordinate Jacobian
\begin{equation}
    \mathcal{J}=\frac{B_\phi+\iota B_\theta}{B^2}, \label{eqn:jacCoord}
\end{equation}
where the covariant $B_\theta$ and $B_\phi$ are flux functions,
\begin{equation*}
    \mathbf{B}=B_\theta(\psi)\nabla\theta+B_\phi(\psi)\nabla\phi+B_\psi\nabla\psi.
\end{equation*}}
Boozer coordinates are widely used in stellarator theory and applications. The coordinates are a natural set that simplify many of the governing equations, including QS. In particular, construction of solutions by near-axis expansion of three-dimensional fields is most convenient in these coordinates\cite{garrenboozer1991a,garrenboozer1991b,landreman2018a,rodriguez2020i,rodriguez2020ii}. \par
In the context of our general quasisymmetric $\mathbf{B}$, it is not necessary to assume that $\mathbf{j}\cdot\nabla\psi=0$. However, we demonstrate that any quasisymmetric field must satisfy $\oint (\mathrm{d}l/B)(\mathbf{j}\cdot\nabla\psi)=0$ (see Appendix A). The question that naturally arises is, can we construct an appropriate straight field line coordinate system that explicitly expresses the symmetry in QS in the Boozer fashion? \par
The answer is yes. To see this, we write Eq.~(\ref{eqn:tripleQS}) in straight-field line coordinates $\{\psi,\theta,\phi\}$ using the contravariant representation $\mathbf{B}=\nabla\psi\times(\nabla\theta-\iota\nabla\phi)$,
\begin{multline*}
    (\nabla\psi\times\nabla\theta\:\partial_\theta B+\nabla\psi\times\nabla\phi\:\partial_\phi B)\cdot\\
    \cdot\nabla(\mathcal{J}^{-1}(\partial_\phi B+\iota \partial_\theta B))=0,
\end{multline*}
where $\mathcal{J}$ is the Jacobian associated to the coordinate system chosen. Now assume we can construct a straight-field line coordinate system with a Jacobian of the form $\mathcal{J}=\mathcal{J}(\psi,B)$, then the Jacobian factor may be dropped from the equation above to obtain,
\begin{gather*}
    (\nabla\psi\times\nabla\theta\:\partial_\theta B+\nabla\psi\times\nabla\phi\:\partial_\phi B)\cdot\nabla(\partial_\phi B+\iota \partial_\theta B)=0 \\
    \text{i.e.}\quad \left[\partial_\phi-\left(\frac{\partial_\phi B}{\partial_\theta B}\right)\partial_\theta\right](\partial_\phi+\iota\partial_\theta)B=0
\end{gather*}
assuming that $\partial_\theta B\neq0$. (This makes quasi-poloidally symmetric solutions a special case.) From near-axis expansion we know that these solutions have a very restricted QS region\cite{rodriguez2020}. \par
Commuting operators, we obtain
\begin{equation*}
    (\partial_\phi+\iota\partial_\theta)\left(\frac{\partial_\phi B}{\partial_\theta B}\right)\partial_\theta B=(\mathbf{b}\cdot\nabla)\left(\frac{\partial_\phi B}{\partial_\theta B}\right)\partial_\theta B=0
\end{equation*}
which implies that $B=B(\psi,\theta-\Tilde{\alpha}\phi)$ or $B=B(\psi,\phi)$, where $\Tilde{\alpha}=-\partial_\phi B/\partial_\theta B$ is a flux function. The additional requirement that $\Tilde{\alpha}$ is rational to avoid $B=B(\psi)$ at non-degenerate surfaces requires $\Tilde{\alpha}$ to be constant.\cite{rodriguez2020}  \par
In summary, if we are able to construct a straight field line coordinate system that has a Jacobian that can be written as a function of $\psi$ and $B$ only, then a field is QS in the \textit{weak} sense if and only if $B$ depends on a single linear combination of those coordinate angles. Note that under this assumption, the reverse direction of the proof is straightforward.

\subsection{Constructing generalized Boozer coordinates} \label{sec:GBCconstr}
We define \textit{generalized Boozer coordinates} (GBC) as a set of straight field line coordinates whose Jacobian can be written in the form $\mathcal{J}=B_\alpha(\psi)/B^2$, where $B_\alpha$ is some flux function  {{}without requiring   $\mathbf{j}\cdot\nabla\psi$ to vanish identically. This choice is more general than what is permitted by Boozer coordinates, which separately requires $B_\theta$ and $B_\phi$ to be flux functions. We shall now constructively explore the conditions under which such a coordinate system exists.} \par
Let us start with a given magnetic field, assuming that it lies on well-defined flux surfaces. It then follows\cite{kruskuls1958,boozer1981,Helander2014} that there exists some initial straight field line coordinate system $\{\psi,\theta,\phi\}$, so that $\mathbf{B}=\nabla\psi\times\nabla\theta+\iota\nabla\phi\times\nabla\psi$. The definition of the angular straight field line coordinates is arbitrary up to a transformation of the form
\begin{gather*}
\theta=\theta'+\iota\omega, \\
\phi=\phi'+\omega.
\end{gather*}
The function $\omega=\omega(\psi,\theta,\phi)$ is a well behaved periodic function that defines a family of transformations\cite{Helander2014}. The periodicity of $\omega$ preserves the toroidal and poloidal character of the two angular coordinates. \par
Starting with some given straight-field line coordinate system, we want to understand how to construct a set with a Jacobian of the GBC form. Thus, we need to know how to transform the coordinate Jacobian induced by $\omega$. This reads
\begin{equation*}
    \mathcal{J}'^{-1}=\nabla\psi\times\nabla(\theta-\iota\omega)\cdot\nabla(\phi-\omega). \label{jacTrans}
\end{equation*}
Here $\{\psi,\theta',\phi'\}$ represents the newly defined straight field line coordinates whose associated Jacobian is $\mathcal{J}'$. This equation may be recast into the form of a magnetic differential equation,
\begin{equation}
    \mathbf{B}\cdot\nabla\omega=\frac{1}{\mathcal{J}}-\frac{1}{\mathcal{J}'}. \label{eqn:magDifEqOr}
\end{equation}
Now, let us require\footnote{A more general form $\mathcal{J}'=\mathcal{J}'(\psi,B)$ could have been demanded. However, one may show in that case that the system may always be cast into the form here employed.}
\begin{equation}
    \mathcal{J}'=\frac{B_\alpha(\psi)}{B^2}. \label{eqn:GBCjac}
\end{equation}
In order for the magnetic differential equation Eq.~(\ref{eqn:magDifEqOr}) to have a single-valued solution for the transformation function $\omega$, Eq.~(\ref{eqn:magDifEqOr}) must satisfy Newcomb's criterion\cite{newcomb1959}. According to Newcomb, for a magnetic differential equation $\mathbf{B}\cdot\nabla f=s$ to have a single-valued solution for $f$, the source term $s$ must satisfy the line-integral condition along a field line,
\begin{equation}
    \oint s\frac{\mathrm{d}l}{B}=0.
\end{equation}
For our problem $s=1/\mathcal{J}-1/\mathcal{J}'$. Start with,
\begin{equation*}
    \oint \frac{1}{\mathcal{J}}\frac{\mathrm{d}l}{B}= \oint\mathbf{B}\cdot\nabla\phi\frac{\mathrm{d}l}{B}=\oint \mathrm{d}\phi = 2\pi n.
\end{equation*}
Here we have used the definition of the Jacobian in terms of the magnetic field $\mathbf{B}\cdot\nabla\phi=\nabla\psi\times\nabla\theta\cdot\nabla\phi=1/\mathcal{J}$, where $\phi$ and $\theta$ are, respectively, toroidal and poloidal angular coordinates that increase by $2\pi$ in going around the long or short toroidal circular paths. We also considered the rotational transform $\iota=n/m$, where $n,m\in\mathbb{Z}$ and are coprime. \par
Now, consider Newcomb's condition on the last term on the right-hand-side of Eq.~(\ref{eqn:magDifEqOr}),
and define the closed line integral $\mathcal{I}$ so that,
\begin{equation}
    \oint \frac{1}{\mathcal{J}'}\frac{\mathrm{d}l}{B}=\frac{\mathcal{I}}{B_\alpha(\psi)}.
\end{equation}
For Newcomb's condition on Eq.~(\ref{eqn:magDifEqOr}) to hold, the following must be true:
\begin{equation}
    B_\alpha(\psi)=\frac{\mathcal{I}}{2\pi n}. \label{eqn:BaDef}
\end{equation}
With this choice
\begin{equation}
    \oint \mathbf{B}\cdot\nabla\omega\frac{\mathrm{d}l}{B}=0.
\end{equation}
Then there must exist a single-valued solution $\omega$ that enacts the coordinate transformation into the set associated with the Jacobian (\ref{eqn:GBCjac}). \par
For Eq.~(\ref{eqn:BaDef}) to hold, it is necessary for $\mathcal{I}$ to be a flux function. This condition may be written in the form,
\begin{equation}
    \mathcal{I}(\psi)=\oint\mathbf{B}\cdot\mathrm{d}\mathbf{r},
\end{equation}
\begin{figure}
    \includegraphics[width=5cm]{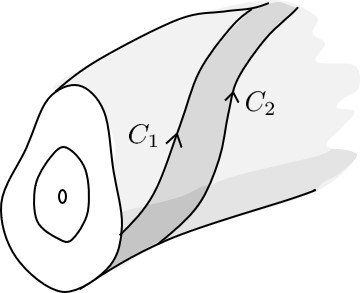}
    \caption{\textbf{Ribbon surface.} Ribbon defined by two closely lying rational magnetic field lines labelled by $C_1$ and $C_2$.}
    \label{fig:ribbonGBC}
\end{figure}
along a magnetic field line. Focus on the case of rational field lines, for which the condition is most stringent. To proceed further, consider a non-self-intersecting ribbon over a flux surface bounded by two adjacent magnetic field lines (see Fig.~\ref{fig:ribbonGBC}). Denote the line integrals along each of these lines by $C_1$ and $C_2$. We may then write, using Stoke's theorem,
\begin{equation}
    \oint_{C_1}\mathbf{B}\cdot\mathrm{d}\mathbf{r}-\oint_{C_2}\mathbf{B}\cdot\mathrm{d}\mathbf{r}=\int_\mathrm{rib}\mathbf{j}\cdot\mathrm{d}\mathbf{S}, \label{eqn:ribbonInt}
\end{equation}
where the surface element is taken to be perpendicular to the flux surface. The integral over the surface may be written as,\cite{Helander2014}
\begin{equation*}
    \int_\mathrm{rib}\mathbf{j}\cdot\mathrm{d}\mathbf{S}=\int_{\alpha_0}^{\alpha_0+\delta\alpha}\mathrm{d}\alpha\oint\frac{\mathrm{d}l}{B}\mathbf{j}\cdot\nabla\psi.
\end{equation*}
Here $\alpha$ labels the field lines on the surface. Now, if $\mathcal{I}$ is to be truly a flux function, then following (\ref{eqn:ribbonInt}) the last surface integral must vanish. And it must do so for all field line labels $\alpha_0$ and adjacent field lines $\delta\alpha$. This gives the necessary and sufficient condition,
\begin{equation}
    \oint\frac{\mathrm{d}l}{B}\mathbf{j}\cdot\nabla\psi=0. \label{eqn:subclassGBC}
\end{equation}
This subclass of magnetic fields grant the required form of $\mathcal{I}$, and thus the single-valued solution to Eq.~(\ref{eqn:magDifEqOr}). This subclass includes the stronger constraint $\mathbf{j}\cdot\nabla\psi=0$ {{} (for which the coordinates reduce to Boozer coordinates)}, or more to our concern, the QS scenario (see Appendix A). Note that magnetic fields that possess nested surfaces have the property $\langle\mathbf{j}\cdot\nabla\psi\rangle=0$, following an application of $\nabla\cdot(\mathbf{B}\times\nabla\psi)$. However, the Newcomb condition (\ref{eqn:subclassGBC}) is more stringent.\cite{newcomb1959} \par
Restricting ourselves to this subclass of fields, we can always choose $B_\alpha$ as in (\ref{eqn:BaDef}). This choice might appear artificial, especially given the presence of the discrete $n$. However, we may relate this to the surface average over irrational flux surfaces. To see this, we write,
\begin{multline}
    B_\alpha=\frac{1}{2\pi n}\oint B^2\frac{\mathrm{d}l}{B}=\frac{1}{4\pi^2 n}\int_0^{2\pi}\mathrm{d}\alpha\underbrace{\oint B^2\frac{\mathrm{d}l}{B}}_{n~\mathrm{turns}}=\\
    =\frac{1}{4\pi^2}\int_0^{2\pi}\mathrm{d}\alpha\underbrace{\oint B^2\frac{\mathrm{d}l}{B}}_{1~\mathrm{turn}}=\frac{V'}{4\pi^2}\langle B^2\rangle, \label{eqn:BaIrrat}
\end{multline}
where $V'=\int_0^{2\pi}\mathrm{d}\alpha\int\mathrm{d}l/B=V'\langle 1\rangle$ is the usual volume $\psi$ derivative. {{} $B_\alpha$ is, therefore, a well behaved quantity for both rational and irrational surfaces.} \par
As it stands, with an appropriate choice of $B_\alpha$, and restricting ourselves to the subclass of fields satisfying Eq.~(\ref{eqn:subclassGBC}), one may perform a coordinate transformation that provides a Jacobian of the desired form (\ref{eqn:GBCjac}). It remains to be shown that this new coordinate system is well-behaved. By this we mean that the Jacobian does not vanish nor diverge. To see this, it is most useful to rewrite $\mathcal{J}'$ in terms of (\ref{eqn:BaIrrat}),
\begin{equation}
    \mathcal{J}'=\frac{B_\alpha}{B^2}=\frac{V'}{4\pi^2}\frac{\langle B^2\rangle}{B^2}.
\end{equation}
It is clear that $\mathcal{J}'$ has a definite sign, as given by the sign of $V'$. Thus, the Jacobian will never vanish nor diverge, given that $B^2>0$.

\subsection{Magnetic field in generalized Boozer coordinates}
We have shown that under the assumption that the magnetic field is quasisymmetric, we have a straight-field-line coordinate system, GBC, with a Jacobian $J=B_\alpha(\psi)/B^2$. In this coordinate system and following Sec.~\ref{sec:GBCQS}, a quasisymmetric field is one whose magnitude can be expressed in GBC as $B=B(\psi,\theta-\Tilde{\alpha}\phi)$. This form is analogous to the Boozer formulation of QS but is only subordinated to the configuration being QS and not $\mathbf{j}\cdot\nabla\psi=0$.  \par
Before proceeding to analyze some of the properties of QS and other governing equations in GBC, we explicitly write the covariant and contravariant forms for $\mathbf{B}$. The covariant form is
\begin{gather}
    \mathbf{B}=B_\theta\nabla\theta+(B_\alpha-\iota B_\theta)\nabla\phi+B_\psi\nabla\psi. \label{eqn:BinGBC}
\end{gather}
The usual covariant function $B_\phi$ has been deprecated for $B_\alpha$. This is the flux function that appears in the Jacobian (\ref{eqn:GBCjac}). The simplicity of (\ref{eqn:BinGBC}) shows why it was convenient to choose the Jacobian of GBC to have the particular form in (\ref{eqn:GBCjac}). To obtain (\ref{eqn:BinGBC}), it is sufficient to take its scalar product with the contravariant representation,
\begin{equation}
    \mathbf{B}=\nabla\psi\times\nabla\theta+\iota\nabla\phi\times\nabla\psi, \label{eqn:contra}
\end{equation}
and capitalize on the definition of $\mathcal{J}$. Compared to Boozer coordinates, the covariant function $B_\theta$ in GBC is not necessarily a flux function. Thus, GBC is an extension of Boozer coordinates. {{} When $\mathbf{j}\cdot\nabla\psi=0$, however, and as previously pointed, GBCs reduce to Boozer coordinates.} So far, the forms in (\ref{eqn:BinGBC}) and (\ref{eqn:contra}) have only required of the existence of GBC, and not of QS per se --other than to guarantee (\ref{eqn:subclassGBC}). To enforce the latter, one needs to specify $|\mathbf{B}|$ as a symmetric function.\cite{rodriguez2020i,rodriguez2020ii,sengupta2021} 

\section{Describing weak quasisymmetry in GBC}
Having developed GBC, let us see what this coordinate system can teach us about \textit{weak} QS. We first write down the complete set of equations that describe a weakly quasisymmetric magnetic field. The first relevant equation equates the covariant (\ref{eqn:BinGBC}) and contravariant (\ref{eqn:contra}) forms of the magnetic field. The equation reads,
\begin{multline}
    B_\theta\nabla\theta+(B_\alpha-\iota B_\theta)\nabla\phi+B_\psi\nabla\psi=\\
    =\nabla\psi\times\nabla\theta+\iota\nabla\phi\times\nabla\psi. \label{eqn:contravariantEq}
\end{multline}
To specify that we are using GBC, and to introduce the quasisymmetric condition, we require
\begin{equation}
    \nabla\psi\times\nabla\theta\cdot\nabla\phi=\frac{B_\alpha(\psi)}{B^2(\psi,\theta-\Tilde{\alpha}\phi)}. \label{eqn:jacobianEq}
\end{equation}
This set of equations (\ref{eqn:contravariantEq}) and (\ref{eqn:jacobianEq}) is entirely self-contained. It describes a {{} general} magnetic field (a vector field satisfying $\nabla\cdot\mathbf{B}=0=\mathbf{B}\cdot\nabla\psi${{}, without a particular form of equilibrium}) that is quasisymmetric ---no more, no less. Such equations have been recently used in near-axis expansions with anisotropic pressure equilibria.\cite{rodriguez2020i,rodriguez2020ii} Equations (\ref{eqn:contravariantEq}) and (\ref{eqn:jacobianEq}), referred to as the \textit{co(ntra)variant} and \textit{Jacobian} equations respectively, were there expanded systematically. A more thorough and complete exploration of the expansion of the magnetic equations and its implications will be presented in a separate publication. \par
Equations~(\ref{eqn:contravariantEq}) and (\ref{eqn:jacobianEq}) apply beyond near-axis expansions. Alternatively, one could study the behavior of this system in a perturbative sense but around a surface rather than the axis.\cite{sengupta2021} This could shed some light to standard optimization approaches to QS\cite{henneberg2019,bader2019}. \par
Beyond the set of equations (\ref{eqn:contravariantEq}) and (\ref{eqn:jacobianEq}), the formulation of weak QS in terms of GBC can be used to compare \textit{weak} QS to other (quasi)symmetric forms. This comparison to \textit{strong} and axisymmetry will help frame the notion of \textit{weak} QS in the larger space of configurations. 

\subsection{Comparison to strong quasisymmetry} \label{sec:compweakstrong}
\textit{Strong} QS is a more resrictive form of QS compared to its \textit{weak} form\cite{burby2019,rodriguez2020}. Weak QS is the necessary and sufficient condition for the leading guiding centre dynamics to have an \textit{approximately} conserved momentum to leading gyro-centre order.\cite{rodriguez2020} This condition can be written as in (\ref{eqn:tripleQS}), but is most naturally given as,
\begin{gather}
    \mathbf{u}\cdot\nabla B=0, \label{eqn:u.dB}\\
    \mathbf{B}\times\mathbf{u}=\nabla\Phi(\psi), \label{eqn:Bxu}\\
    \nabla\cdot\mathbf{u}=0.\label{eqn:d.u}
\end{gather}
The vector field $\mathbf{u}$ is defined by these equations and points in the direction of symmetry. In \textit{strong} QS the conserved momentum for the particle dynamics is exact for the first-order guiding centre Lagrangian.\cite{burby2019,rodriguez2020,tessarotto1996} In the notation that explicitly introduces the symmetry vector $\mathbf{u}$, strong QS is equivalent to weak QS -- i.e., Eqs.~(\ref{eqn:u.dB})-(\ref{eqn:d.u})-- plus the constraint 
\begin{equation}
    \mathbf{j}\times\mathbf{u}+\nabla C=0, \label{eqn:strongCond}
\end{equation}
where $C=\mathbf{B}\cdot\mathbf{u}$. Note that only the $\mathbf{b}$ component of this additional constraint is contained in the weak formulation of the problem. We now explore the significance of (\ref{eqn:strongCond}) in the context of GBC. \par
To begin, we need to construct $\mathbf{u}$. From Eqs.~(\ref{eqn:u.dB})-(\ref{eqn:d.u}),\cite{rodriguez2020}
\begin{align}
    \mathbf{u}=\Bar{\iota}\frac{\mathbf{\nabla}\psi\times \mathbf{\nabla}\chi}{\mathbf{B}\cdot \mathbf{\nabla}\chi},
    \label{eqn:u_def}
\end{align}
where $\chi=\theta-\Tilde{\alpha}\phi$, and it is convenient to use $\chi$ as part of the coordinate triplet $\{\psi,\chi,\phi\}$. We have made the choice $\Phi'=\Bar{\iota}$ so that $\mathbf{u}\cdot\nabla\phi=1$, for simplicity. Scaling of the flux label for $\mathbf{u}$ leaves the weak QS conditions unchanged (see Appendix B for further discussion on the gauge and the particular choice). This form of $\mathbf{u}$ is equivalent to Eqs.~(\ref{eqn:u.dB})-(\ref{eqn:d.u}) only if we enforce $B=B(\psi,\chi)$ and the coorrdinate system $\mathbf{B}\cdot\nabla\chi=\Bar{\iota}\mathcal{J}^{-1}$ with the Jacobian in (\ref{eqn:GBCjac}). The parameter $\Bar{\iota}=\iota-\Tilde{\alpha}$ has been defined.\par
From (\ref{eqn:BinGBC}) and (\ref{eqn:u_def}), we find,
\begin{equation}
    C=B_\alpha-\Bar{\iota}B_\theta. \label{eqn:CandBtheta}
\end{equation}
This relation (\ref{eqn:CandBtheta}), together with (\ref{eqn:u_def}) and the definition of $C$ can be used to write the gauge-independent form,
\begin{equation}
    \mathbf{B}\cdot\nabla\psi\times\nabla B= \frac{B_\alpha-\Bar{\iota}B_\theta}{\Bar{\iota}}\mathbf{B}\cdot\nabla B. \label{eqn:prePaulForm}
\end{equation}
The magnetohydrostatic form of this equation has been used previously\cite{paul2020,Helander2014}. \par
The contravariant form of $\mathbf{u}$ together with (\ref{eqn:contra}) give simple expressions for directional derivatives in GBC. The differential operators can be simply written as partial derivatives,
\begin{align}
    \mathbf{B}\cdot\mathbf{\nabla}= \mathcal{J}^{-1}\left(\Bar{\iota}\partial_\chi +\partial_\phi\right),\quad \mathbf{u}\cdot\mathbf{\nabla}=\partial_\phi.
    \label{eqn:BDel_uDel}
\end{align}
Since the Jacobian is quasisymmetric from (\ref{eqn:GBCjac}), the two operators $\mathbf{B}\cdot\mathbf{\nabla}$ and $\mathbf{u}\cdot\mathbf{\nabla}$ commute with each other\cite{burby2021}. This commutation property is made manifest in the GBC. \par
The symmetry field $\mathbf{u}$ can also be written in the following covariant form:
\begin{align}
    \mathbf{u}= u_\psi \mathbf{\nabla}\psi +u_\chi \mathbf{\nabla}\chi +u_\phi \mathbf{\nabla}\phi.
    \label{eqn:u_deaf_covid}
\end{align}
Taking the scalar product with $\mathbf{u}$ and $\mathbf{B}$ we obtain
\begin{align}
    u_\phi =u^2 ,\quad \Bar{\iota}u_\chi + u_\phi = \frac{C}{\mathcal{J}^{-1}}.
    \label{eqn:u_covid_symptoms}
\end{align}
 
 To complete the family of vectors required for the strong quasisymmetric condition (\ref{eqn:strongCond}), we need a closed form for $\mathbf{j}$ in GBC. From the curl of the covariant form of $\mathbf{B}$ in Eq.~(\ref{eqn:BinGBC}), we obtain
\begin{equation}
\mathbf{j}=\mathbf{\nabla}B_\psi\times \mathbf{\nabla}\psi + \mathbf{\nabla}B_\theta\times \mathbf{\nabla}\chi +\mathbf{\nabla}(B_\alpha-\Bar{\iota} B_\theta)\times \mathbf{\nabla}\phi.
\label{eqn:current_affairs}
\end{equation}
Using \eqref{eqn:current_affairs} and (\ref{eqn:CandBtheta}), we can show that
\begin{multline}
    \mathbf{j}\times  \mathbf{u}+\mathbf{\nabla} C = (\mathbf{u}\cdot\mathbf{\nabla}B_\psi) \mathbf{\nabla}\psi+(\mathbf{u}\cdot\mathbf{\nabla} B_\theta)\left( \mathbf{\nabla}\chi - \Bar{\iota}\mathbf{\nabla}\phi\right).
    \label{eqn:Jxu}
\end{multline}
The simplicity of (\ref{eqn:Jxu}) is due to the choice of GBC. \par
Recall that strong QS requires the expression $ \mathbf{j}\times  \mathbf{u}+\mathbf{\nabla} C$ to be identically zero. This means that all the terms on the right side of \eqref{eqn:Jxu} need to vanish. That is to say, the covariant functions $(B_\psi, B_\theta)$ are required to be quasisymmetric. If $B_{\theta}$ is quasisymmetric, then $C$ is automatically so from \eqref{eqn:CandBtheta}. In an explicit coordinate representation, using (\ref{eqn:BDel_uDel}), we may write $B_\theta(\psi,\chi)$ and $B_\psi(\psi,\chi)$. \par
Thus, the GBC representation provides an elegant way to formulate strong QS, which can now be understood as weak quasisymmetry plus the conditions that $B_\psi$ and $B_\theta$ are QS. In other words, not only is $B$ QS but so are $B_\theta$ and $B_\psi$. \par
\paragraph{Implications for near-axis expansion}. We refer to [\onlinecite{rodriguez2020i}] and [\onlinecite{rodriguez2020ii}] for a detailed treatment using near-axis expansions of the weakly quasisymmetric problem. The procedure is based on expanding all governing equations describing the weak quasisymmetric field in some form of equilibrium in powers of the distance from the magnetic axis. To do so efficiently, both the field and equations --including Eqs.~(\ref{eqn:contravariantEq}) and (\ref{eqn:jacobianEq})-- are expressed in GBC. It was shown there that when equilibria with anisotropic pressure are considered, the common overdetermination problem\cite{garrenboozer1991b,landreman2018a} that limits the expansion is overcome. The number of governing equations becomes the same as that of functions to be solved. \par
Extending the expansion to strong QS is straightforward. From the discussion above, the only difference is that the covariant functions $B_\theta$ and $B_\psi$ are both quasisymmetric rather than general functions of space. In practical terms, this simply leads to more restricted Taylor-Fourier expansions of those functions; the coefficients that were functions of $\phi$ become constants. This restriction in freedom once again leads us back to the \textit{Garren-Boozer overdetermination} problem. In fact, it does so the same way as it did in the case of MHS equilibrium. The restriction on the covariant functions imposes very severe constraints on the allowed geometry. The only way to escape this impasse is to assume axisymmetry ($\phi$ independent). Once again, consistent with what we observed in [\onlinecite{rodriguez2020i}], the asymmetry of the covariant representation of $\mathbf{B}$ appears to be vital to the construction of QS solutions. \par

\subsection{Comparison to axisymmetry}
We have seen that the strong formulation of QS is more constraining than its weak form. We would also like to compare QS to the limiting case of axisymmetry. We shall think of this case as a symmetry generated by rotation in space: the system is invariant under rotations about an axis. In Euclidean space, a rotation is an isometry, and it is generated by a vector field known as a Killing vector. Using the notion of a Killing vector, we want to explore how `far' the weak concept of QS is from this `true' symmetry. \par
A measure of the departure of a symmetry generator from a Killing vector is the so-called deformation metric.\cite{burby2019} Taking $\mathbf{u}$ to represent the symmetry vector for QS, the idea is to see how far it is from being a Killing vector. A vector field $\mathbf{v}$ is Killing if and only if the deformation tensor $\mathcal{L}_{\mathbf{v}}g=0$. Here $\mathcal{L}_u$ denotes the Lie derivative along $\mathbf{u}$ and $g$ is the Euclidean metric. In 3D, this may be written as,
\begin{equation}
    \mathcal{L}_{\mathbf{u}}g=\nabla\mathbf{u}+(\nabla\mathbf{u})^T.
\end{equation}
Evaluating this tensor for a quasisymmetric configuration should then provide information regarding the closeness to an isometry. It is convenient\cite{burby2019} to evaluate this rank-2 tensor in a basis defined by $\{\mathbf{B},\mathbf{u},\nabla\psi\}$, a triplet that we shall take to be independent. Then,
\begin{align}
    \left[\nabla\mathbf{u}+(\nabla\mathbf{u})^T\right]\cdot\mathbf{B}=&\mathbf{j}\times\mathbf{u}+\nabla C, \label{eqn:Lg1}\\
    \left[\nabla\mathbf{u}+(\nabla\mathbf{u})^T\right]\cdot\mathbf{u}=&\nabla u^2-\mathbf{u}\times\nabla\times\mathbf{u}, \label{eqn:wDef}\\
    \left[\nabla\mathbf{u}+(\nabla\mathbf{u})^T\right]\cdot\nabla\psi=&\nabla\psi\times\nabla\times\mathbf{u}+2\nabla\psi\cdot\nabla\mathbf{u}, \label{eqn:Lg3}
\end{align}
where the \textit{weak} quasisymmetric properties have been used where necessary. Equation~(\ref{eqn:wDef}) is what Burby et al. call $\mathbf{w}$.\cite{burby2021} We shall explore this vector $\mathbf{w}$ in more detail after obtaining an explicit form for $\mathcal{L}_{\mathbf{u}}g$. Using (\ref{eqn:Lg1})-(\ref{eqn:Lg3}), and projecting once again onto the non-orthogonal basis triplet, we obtain
\begin{equation}
    \mathcal{L}_{\mathbf{u}}g=\begin{pmatrix}
    0 & \mathbf{u}\cdot\nabla C & \nabla\psi\cdot(\mathbf{j}\times\mathbf{u}+\nabla C) \\
    \dots & \mathbf{u}\cdot\mathbf{w} & \mathbf{w}\cdot\nabla\psi \\
     \dots & \dots &  -\mathbf{u}\cdot\nabla|\nabla\psi|^2
    \end{pmatrix}.
\end{equation}
The matrix is symmetric by construction. The content of its elements can be made clearer using GBC explicitly. \par
The top row, corresponding to (\ref{eqn:Lg1}), has already been dealt with, as it is precisely the piece corresponding to strong QS. We made the observation that for this condition to be satisfied, (\ref{eqn:Jxu}) required $\mathbf{u}\cdot\nabla B_\theta=0=\mathbf{u}\cdot\nabla B_\psi$. \par
For the other components, the vector field $\mathbf{w}= \mathbf{\omega_u}\times \mathbf{u} + \mathbb{\nabla} u^2, \quad \mathbf{\omega_u}=\mathbf{\nabla}\times \mathbf{u}$ is key. Using the covariant form of $\mathbf{u}$ we obtain the curl of the vector $\mathbf{u}$ in the form
\begin{align}
\mathbf{\omega_u}=\mathbf{\nabla}u_\psi\times \mathbf{\nabla}\psi + \mathbf{\nabla}u_\chi\times \mathbf{\nabla}\chi +\mathbf{\nabla}u_\phi\times \mathbf{\nabla}\phi.
\label{eqn:omega_u_def}
\end{align}
Taking the cross product with $\mathbf{u}$, using the orthogonality of $\mathbf{u}$ with $\mathbf{\nabla}\psi$ and $ \mathbf{\nabla}\chi$, and (\ref{eqn:CandBtheta}) and (\ref{eqn:u_covid_symptoms}) we get
\begin{multline}
    \mathbf{w}
    = (\mathbf{u}\cdot\mathbf{\nabla}u_\psi)\mathbf{\nabla}\psi + (\mathbf{u}\cdot\mathbf{\nabla}u^2)\left(\mathbf{\nabla}\phi - \frac{1}{\Bar{\iota}}\mathbf{\nabla}\chi\right)+\\
    -\mathcal{J}(\mathbf{u}\cdot\mathbf{\nabla}B_\theta)\mathbf{\nabla}\chi,
    \label{eqn:w_useful_expression}
\end{multline}
which implies that
\begin{align}
    \mathbf{B}\cdot \mathbf{w}=& -\Bar{\iota}\mathbf{u}\cdot\mathbf{\nabla}B_\theta, 
    \label{eqn:BDotw_expression}\\
    \mathbf{u}\cdot\mathbf{w}=&\mathbf{u}\cdot\nabla u^2.
\end{align}
Most importantly, a vanishing $\mathbf{w}$ implies that the covariant components of the symmetry vector as well as $B_\theta$ are quasisymmetric. \par
To complete the simplification of the metric tensor, we invoke $B^2=(C^2+|\nabla\psi|^2)/u^2$, which follows from the definition of $\mathbf{u}$. This means,
\begin{equation}
    \mathbf{u}\cdot\nabla|\nabla\psi|^2=B^2\mathbf{u}\cdot\nabla u^2-\mathbf{u}\cdot\nabla C^2.
\end{equation}
With this coordinate representation, the dependence of the various metric pieces is made explicit. We may then schematically present the dependence of $\mathcal{L}_{\mathbf{u}}g$ as follows,
\begin{equation}
    \mathcal{L}_{\mathbf{u}}g\sim\begin{pmatrix}
    0 & \boxed{\partial_\phi B_\theta} & \boxed{\substack{\partial_\phi B_\theta,~\partial_\phi B_\psi}} \\
    \dots & \boxed{\partial_\phi u^2} & \boxed{\substack{\partial_\phi u^2,~\partial_\phi  u_\psi\\\partial_\phi B_\theta}} \\
     \dots & \dots &  \boxed{\substack{\partial_\phi B_\theta,~\partial_\phi u^2}}
    \end{pmatrix}. \label{eqn:LugSim}
\end{equation}
The boxed expressions are meant to indicate that the corresponding tensor component vanishes if the expressions there do. If the tensor (\ref{eqn:LugSim}) is to vanish, then the symmetry vector would correspond to a rotation. This is not surprising if one looks at what it means for the components in (\ref{eqn:LugSim}) to vanish. Axisymmetry is reached when the covariant components of the magnetic field and the symmetry vector are themselves symmetric. The latter is intimately related to the geometry, as we may see when writing $\mathbf{u}\propto\partial_\phi \mathbf{x}|_{\chi,\psi}$. \par
From (\ref{eqn:LugSim}), it follows that in some sense, weak QS is far from being an isometry. This is so because only one of the components of the tensor exactly vanishes. The $\phi$ dependence of the functions $B_\psi$, $B_\theta$, $u^2$, and $u_\psi$ takes the configuration away from axisymmetry. These apparent four degrees of freedom (especially those involving $\mathbf{u}$) may not be independent and involve highly non-linear combinations---they should ultimately be related through the quasisymmetric magnetic equations. Anyhow, the field-line dependence is key in distinguishing the weakly quasisymmetric form from, say, an axisymmetric tokamak. \par
To make a comparative measurement of the departure from axisymmetry, consider now the case of strong QS. In this case, following (\ref{eqn:strongCond}), the first whole row (and thus also column) of (\ref{eqn:LugSim}) drop. The remaining dependence also simplifies, and the system is precluded from being an isometry, a priori, through the $\phi$ dependence of $u^2$ and $u_\psi$ only, which is consistent with the work by Burby et al.\cite{burby2019} \par
Imposing additional properties to the field may also affect the form of the deformation tensor. An example would be a particular form of force balance. We now explore how the magnetics and equilibria are linked. \par


\section{Quasisymmetry and equilibria}
Let us consider the force balance part of the problem. Generally, a magnetic equilibrium with some arbitrary force $\mathbf{F}$ reads,
\begin{equation}
    \mathbf{j}\times\mathbf{B}=\mathbf{F}. \label{eqn:Fbal}
\end{equation}
{{} As we argued in Sec.~II, we are concerned in this work with a general fluid force $\mathbf{F}$. Its connection to the microphysics is not considered.} Let us express the left-hand side of (\ref{eqn:Fbal}) in GBC. Using the contravariant form for the current (\ref{eqn:current_affairs}) together with (\ref{eqn:CandBtheta}), we obtain
\begin{multline}
     \mathbf{j}\times  \mathbf{B} = \left[\mathbf{B}\cdot\mathbf{\nabla}B_\psi-\mathcal{J}^{-1}\left( B_\alpha' -\Bar{\iota}' B_\theta\right)\right] \mathbf{\nabla}\psi+\\
     +(\mathbf{B}\cdot\mathbf{\nabla} B_\theta) \left( \mathbf{\nabla}\chi -\Bar{\iota} \mathbf{\nabla}\phi\right),
    \label{eqn:JxB_simplified}
\end{multline}
which is an explicit coordinate representation of $\mathbf{j}\times\mathbf{B}$. The form of (\ref{eqn:JxB_simplified}) mirrors the form of (\ref{eqn:Jxu}). In this case, the magnetic differential operators substitute the directional derivatives along $\mathbf{u}$.  We note that \eqref{eqn:JxB_simplified} does not have any component along $\mathbf{B}$, as can be checked by taking the dot product with $\mathbf{B}$. \par
The form of (\ref{eqn:JxB_simplified}) puts constraints on the allowable forms for $\mathbf{F}$. As already noted, $\mathbf{F}\cdot\mathbf{B}=0$ must hold true. Otherwise the system would be imbalanced, as the magnetic field is unable to exerts forces along $\mathbf{B}$. Because of this reduction in the dimensionality of $\mathbf{F}$ and in view of (\ref{eqn:JxB_simplified}), it is convenient to write 
\begin{align}
\mathbf{F}=F_\psi \nabla \psi + F_\alpha \left(\nabla \chi - \Bar{\iota}\nabla \phi\right).
    \label{eq:F_form}
\end{align}
An alternative form would be to use the contravariant form of $\mathbf{B}$ \eqref{eqn:contra} in \eqref{eqn:Fbal} to get
\begin{align}
    \mathbf{F}=  \left(\mathbf{J\cdot\nabla}\chi-\Bar{\iota}\mathbf{J\cdot\nabla}\phi \right)\nabla \psi -(\mathbf{J\cdot\nabla}\psi) \left(\nabla \chi - \Bar{\iota}\nabla \phi\right).
    \label{eq:F_n_jdel}
\end{align}
Substituting \eqref{eqn:JxB_simplified} and \eqref{eq:F_form} into \eqref{eqn:Fbal} we get two magnetic differential equations
\begin{subequations}
\begin{align}
   &\mathbf{B}\cdot\mathbf{\nabla}B_\psi=F_\psi + \mathcal{J}^{-1}\left(B_\alpha'-\Bar{\iota}' B_\theta\right), \label{eqn:MDE_B_psi}\\ 
   &\mathbf{B}\cdot\mathbf{\nabla} B_\theta=F_\alpha= -\mathbf{J\cdot \nabla}\psi. \label{eqn:MDE_B_theta}
\end{align}
\label{eqn:MDE_B_psi_B_theta}
\end{subequations}
Therefore, the generalized force-balance condition is equivalent to two magnetic differential equations (MDEs) and $\mathbf{B\cdot F}=0$. If solutions to these equations can be found together with the magnetic equations (\ref{eqn:contravariantEq}) and (\ref{eqn:jacobianEq}), we will have obtained a quasisymmetric configuration in equilibrium. \par
Let us describe in more detail the implications of these equations. First look at the simpler (\ref{eqn:MDE_B_theta}). This equation has two pieces to it. First, and regardless of the assumed form for $F_\alpha$, it follows from weak QS that $\mathbf{B}\cdot\nabla B_\theta=-\mathbf{j}\cdot\nabla\psi$ (see Appendix A as well). This imposes the condition (\ref{sec:GBCconstr}) on the field. Secondly, the component $F_\alpha$ of the forcing $\mathbf{F}$ directly sets the off-surface current. This means that,\cite{newcomb1959} 
\begin{equation}
    \oint \frac{d\ell}{B} F_\alpha =0.
    \label{eqn:Newcomb_F_alpha}
\end{equation}
We may not choose $F_\alpha$ arbitrarily. It must satisfy (\ref{eqn:Newcomb_F_alpha}) if the force is to be consistent with QS --condition like (\ref{eqn:subclassGBC}). Then the magnetic differential equation can be satisfied, and one can directly relate $B_\theta$ and $F_\alpha$ up to a flux function. In Fourier representation, it is clear that the $\phi$ content of $B_\theta$ will be non-zero only if that of $F_\alpha$ is (and vice versa). In the light of (\ref{eqn:LugSim}), choosing $\partial_\phi(\mathbf{j}\cdot\nabla\psi)=0$ brings the quasisymmetric configuration closer to an isometry. This freedom in the form of $B_\theta$ does not exist in the strong formulation of QS, for which $\mathbf{j}\cdot\nabla\psi$ is independent of the field line label. \par
A similar analysis is suitable for (\ref{eqn:MDE_B_psi}). The appropriate Newcomb condition in this case is,
\begin{align}
    \oint \frac{d\ell}{B} \left[F_\psi + \mathcal{J}^{-1}\left(B_\alpha'-\Bar{\iota}' B_\theta\right) \right]=0.
    \label{eqn:Newcomb_F_psi}
\end{align}
This condition may be understood as an averaged radial equilibrium equation. A similar solvability condition can be found, for the special case of MHS, in [\onlinecite{tessarotto1995}], where the notion of a QS field is presented as one that satisfies the Newcomb conditions. Given Eq.~(\ref{eqn:Newcomb_F_psi}), Eq.~(\ref{eqn:MDE_B_psi}) relates $B_\psi,~B_\theta$ (or $F_\alpha$) and $F_\psi$. Once again, we see the close relationship between the forcing, the magnetic covariant representation, and the deviation from axisymmetry. A $\phi$ dependence on $B_\psi$ will bring a finite deviation of $\mathbf{u}$ from being a Killing vector. However, it will also force $F_\alpha$ and $F_\psi$ to have a $\phi$ dependence which may require very particular shaping of the forces. This observation is consistent with the Constantin-Drivas-Ginsberg theorem\cite{constantin2021}, in which the forcing is seen to be intimately related to the deviation from an isometry. Here the asymmetric geometry, quasisymmetry, and the forcing are all intimately connected.   \par
When the magnetic differential equations imposing force balance are brought together with the magnetic equations from the previous section, it is not obvious how the system of equations is to be interpreted: what is to be taken as an input and what should be solved for. Just as an analogy, in the Grad-Shafranov equation, it is clear that $p$ and $F$ are inputs, and $\psi$ is the output. In the present problem, we have the construction of GBC in addition to the various magnetic covariant forms and the components of $\mathbf{F}$. Motivated by the treatment in [\onlinecite{rodriguez2020}] (which deals with a special case of the above), we propose as a possibility for a well-posed problem (\ref{eqn:MDE_B_theta}) to be solved for $B_\theta$ given $F_\alpha$, while $F_\psi$ is the output of (\ref{eqn:MDE_B_psi}), with the function $B_\psi$ specified from the magnetic equations. It is not a trivial matter to determine a convenient way in which to formulate the problem. A more elaborate discussion on this procedure and its implications on constructing solutions is left to future work. 

\subsection{Ideal MHD: $\mathbf{j}\times  \mathbf{B}
=\mathbf{\nabla}p(\psi)$}
\label{sec:ideal_MHD_static}
Let us now revisit the limit of ideal MHD without flows,  $\mathbf{j}\times  \mathbf{B}=\mathbf{\nabla} p$. More general forms will be discussed elsewhere, together with a more systematic treatment of the quasisymmetric system of equations. \par
In ideal MHD, from $\mathbf{j}\cdot\nabla p(\psi)=0$ it follows that 
\begin{gather}
    \mathbf{B}\cdot\mathbf{\nabla} B_\theta=0.
\end{gather}
Thus taking $F_\alpha=0$ forces $B_\theta$ into a flux function. Of course, this also means that $\mathbf{u}\cdot\mathbf{\nabla} B_\theta=0$. Furthermore, as $F_\psi=p'$,  in static ideal MHD, \eqref{eqn:JxB_simplified} leads to the magnetic differential equation for $B_\psi$,
\begin{align}
    \mathbf{B}\cdot\mathbf{\nabla} B_\psi = p'(\psi) + \mathcal{J}^{-1}\left(B_\alpha'-\Bar{\iota}' B_\theta\right)
    \label{eqn:MDE_Bpsi_MHD}
\end{align}
Since $B_\psi$ must be a single-valued function the flux-surface average of \eqref{eqn:MDE_Bpsi_MHD} gives
\begin{align}
    p'(\psi) +\frac{\langle B^2\rangle}{B_\alpha}\left(B_\alpha'-\Bar{\iota}' B_\theta\right)=0.
    \label{eqn:iota_from_MDE_Bpsi}
\end{align}
If we choose the forms of $B$, $p$, and $\iota$, this pins the form of $B_\alpha$ down. Now, looking back to \eqref{eqn:MDE_Bpsi_MHD}, every term on the right-hand side is quasisymmetric. Therefore, $B_\psi$ must also be quasisymmetric if it is to satisfy the force-balance equation. Note that we have already recognized this constraining requirement on the form of $B_\psi$ as the origin of the Garren-Boozer overdetermination problem \cite{rodriguez2020i}. The Newcomb condition on this equation can be recognized as the condition to avoid Pfirsch-Schl\"{u}ter current singularities. \par
The simplifications due to ideal static MHD leads to the vanishing of \eqref{eqn:Jxu}. Therefore, in this limit weak QS is identical to strong QS. \par
One can further show using \eqref{eqn:current_affairs}, \eqref{eqn:BinGBC}, \eqref{eqn:u_def} and \eqref{eqn:MDE_Bpsi_MHD},
\begin{align}
    \mathbf{j}= -\frac{1}{\Bar{\iota}}\partial_\psi (B_\alpha-\Bar{\iota}B_\theta)\mathbf{B}-\frac{1}{\Bar{\iota}}p'(\psi) \mathbf{u},
    \label{eq:JB_JQ_def}
\end{align}
where the gauge choice for $\mathbf{u}$ has been made $\Phi'=\Bar{\iota}$. The expression for $\mathbf{j}$ ought to be $\mathbf{u}$-gauge independent, as it is a physical quantity. The $\mathbf{B}$ piece as written is gauge independent, but the $\mathbf{u}$ term is not. The $\Bar{\iota}$ factor in the latter is to be interpreted as $\Phi'$. This equation had been obtained previously\cite{burby2019,burby2021} using coordinate free, differential forms (see Appendix C). Two special cases of ideal MHD, a) vacuum ($\mathbf{j}=0$) and b) force-free ($\mathbf{j}=\lambda(\psi,\alpha)\mathbf{B} $), are worth pointing to for their importance in plasma physics. For both these cases $p'(\psi)=0$. From \eqref{eq:JB_JQ_def} we see that for the magnetic field to be curl-free (vacuum) and quasisymmetric $C'(\psi)=0$; i.e., $C(\psi)$ must be a constant. For quasisymmetric force-free fields, we must have $\lambda= -C'(\psi)$.
Note that in strong QS these conclusions follow directly from the equation $\mathbf{j\times u}+\mathbf{\nabla}C=0$ with $\mathbf{j}=0$ and $\mathbf{j}=\lambda \mathbf{B}$.

\section{Conclusions}
In this paper, we have presented, defined, and discussed a straight field line coordinate system which is natural for the representation of general-equilibria quasisymmetric magnetic fields: \textit{generalized Boozer coordinates}. We proved the existence of the said coordinate system for the subset of fields for which $\oint\mathbf{j}\cdot\nabla\psi\mathrm{d}l/B=0${{}, to which quasisymmetric fields belong. These coordinates reduce to Boozer coordinates when $\mathbf{j}\cdot\nabla\psi=0$. } \par
The explicit form of the symmetry in this coordinate representation enables a simple formulation of the quasisymmetric problem. We explicitly construct the governing equations setting clearly the foundation for future investigations, including expansion\cite{rodriguez2020i,rodriguez2020ii} and global approaches. Exploiting GBC, we explicitly show the essential differences between the weak and strong formulations of QS and between quasisymmetry and axisymmetry. Weak QS generally lies far from axisymmetry, for which many functions describing the field and symmetry need to be symmetric.  \par 
We also included a set of simple magnetic differential equations that fully describe equilibrium with an arbitrary macroscopic force to complete the treatment. The property of QS, together with the force-balance structure, imposes requirements on the forcing terms in the form of Newcomb conditions. In addition, the equations establish clear connections between QS, forcing, and departures from axisymmetry.

\hfill 
\section*{Acknowledgements}
We are grateful to J. Burby, N. Duigan, J. Meiss, and D. Ginsburg for stimulating discussions. This research is supported by grants from the Simons Foundation/SFARI (560651, AB) and DoE Contract No DE-AC02-09CH11466. 

\section*{Data availability}
Data sharing is not applicable to this article as no new data were created or analyzed in this study.

\appendix
\section{Off-surface current and QS}
In this appendix we directly show how the triple vector formulation of QS in Eq.~(\ref{eqn:tripleQS}) imposes a constraint on the off-normal component of the current at every magnetic surface. \par
Let us start by noting that $\mathbf{B}\cdot\nabla B=f(\psi,B)$ is a consequence of (\ref{eqn:tripleQS}). In that sense, we can reshape the triple vector condition in the convenient form,
\begin{equation}
    \nabla\psi\times\nabla B\cdot\nabla\left(\frac{B^2}{\mathbf{B}\cdot\nabla B}\right)=0. \label{eqn:A1}
\end{equation}
(For now we shall not worry about $\mathbf{B}\cdot\nabla B=0$.) We may express this as a divergence as $\nabla\cdot(\nabla\psi\times\nabla B)=0$. Separating the argument of the divergence into a component parallel to the magnetic field and perpendicular to it, we may rewrite (\ref{eqn:A1}),
\begin{equation}
    \nabla\cdot\left[\frac{\nabla\psi\times\nabla B\cdot\mathbf{B}}{\mathbf{B}\cdot\nabla B}\mathbf{B}+\nabla\psi\times\mathbf{B}\right]=0.
\end{equation}
It is customary to define $C=\nabla\psi\times\nabla B\cdot\mathbf{B}/(\mathbf{B}\cdot\nabla B)$. This can therefore be rewritten, using $\nabla\cdot\mathbf{B}=0$,
\begin{equation}
    \mathbf{B}\cdot\nabla C=\mathbf{j}\cdot\nabla\psi.
\end{equation}
It is then clear that the Newcomb condition\cite{newcomb1959},
\begin{equation}
    \oint\mathbf{j}\cdot\nabla\psi\frac{\mathrm{d}l}{B}=0, \label{eqn:A4}
\end{equation}
with the line integral taken along closed magnetic field lines. For irrational field lines, this amounts to $\langle \mathbf{j}\cdot\nabla\psi\rangle=0$, which is true of all magnetic fields with flux surfaces $\mathbf{B}\cdot\nabla\psi=0$. The condition (\ref{eqn:A4}) is a more constraining condition than the flux surface average one\cite{newcomb1959}.  \par
The division by $\mathbf{B}\cdot\nabla B$ seems to be ill-defined at the extrema of the magnetic field along the field lines. However, the above holds, first, arbitrarily close to these extrema, and thus would expect to hold by continuity. The fact that the condition holds can be seen by looking at the fundamental formulation of QS.\cite{rodriguez2020} As for the behavior of $C$, this is a physical quantity related to the width of banana orbits of bouncing particles. If deeply trapped and barely trapped particles are to be kept, then physically $C$ must be finite at the extrema. Once we express everything in terms of GBC, the apparent divergence of $C$ disappears as both the numerator and denominator are proportional to $\partial_\chi B=0$ when $\mathbf{B}\cdot\nabla B=0$.

\section{Gauge freedom in weak $\mathbf{u}$}
As noted, the definition of $\mathbf{u}$ by the weak quasisymmetry equations (\ref{eqn:u.dB})-(\ref{eqn:d.u}) is invariant to the relabelling of $\Phi$. A rescaling of the flux surface label together with $\mathbf{u}$ leaves the equations invariant, and should therefore have no physical implication in the description of quasisymmetry. Keeping the general gauge label $\Phi(\psi)$ of (\ref{eqn:Bxu}), Eq.~(\ref{eqn:u_def}) reads,
\begin{equation}
    \mathbf{u}=\frac{\nabla\Phi\times\nabla\chi}{\mathbf{B}\cdot\nabla\chi}.
\end{equation}
Then it folllows,
\begin{equation}
    C=\frac{\Phi'}{\Bar{\iota}}(B_\alpha-\Bar{\iota}B_\theta).
\end{equation}
and
\begin{equation}
    \mathbf{u}\cdot\nabla=\frac{\Phi'}{\Bar{\iota}}\partial_\phi.
\end{equation}
In Sec.~\ref{sec:compweakstrong} we then went on to evaluate how the symmetry vector $\mathbf{u}$ as defined by weak QS compares to strong QS. To do so, we needed to evaluate (\ref{eqn:Jxu}), the third of the strong QS conditions. The first thing to note is that this equation is not gauge-invariant. This is, of course, an added complication to the problem. In a sense, this gauge-symmetry breaking leaves no unique form for $\mathbf{j}\times\mathbf{u}+\nabla C$, but rather a whole family. \par
This can be seen explicitly if instead of restricting ourselves to the special case $\Phi'=\Bar{\iota}$, we keep $\Phi$ general. Then, (\ref{eqn:Jxu}) can be written in the form,
\begin{multline}
    \mathbf{j}\times  \mathbf{u}+\mathbf{\nabla} C = \left[\mathbf{u}\cdot\mathbf{\nabla}B_\psi-C\frac{\Phi'}{\Bar{\iota}}\partial_\psi\left(\frac{\Bar{\iota}}{\Phi'}\right)\right] \mathbf{\nabla}\psi+\\
    +(\mathbf{u}\cdot\mathbf{\nabla} B_\theta)\left( \mathbf{\nabla}\chi - \Bar{\iota}\mathbf{\nabla}\phi\right).
    \label{eqn:Jxu_2}
\end{multline}
So the question may be reformulated: what are the conditions for this expression to vanish? As before one piece is $\mathbf{u}\cdot\nabla B_\theta=0$. Now, the other component has the form of a differential equation along the direction of symmetry. We may write it explicitly as,
\begin{equation}
    \partial_\phi B_\psi-(B_\alpha-\Bar{\iota}B_\theta)\partial_\psi\ln(\Bar{\iota}/\Phi')=0.
\end{equation}
For the equation to be consistent it must then hold that either $\Bar{\iota}=a\Phi'$ where $a$ is a constant, or $\int(B_\alpha-\Bar{\iota}B_\theta)\mathrm{d}\phi=0$. The latter is a condition on the physically relevant part of the problem. Thus, it is minimal to choose $\Phi'=\Bar{\iota}$ to satisfy this condition. Then we are simply left with $\mathbf{u}\cdot\nabla B_\psi=0$, which is the choice made in the main text. 

\section{Coordinate independent form of quasisymmetric current}
In this appendix we present a brief derivation of (\ref{eq:JB_JQ_def}) without using coordinates. Start from,
\begin{equation}
    \n{B}\times\n{u}=\nabla\Phi \label{QSBC2}
\end{equation}
with $\Phi$ some single-valued function. With magnetic flux surfaces labelled by $\Phi$, to which $\n{B}$ as well as the vector field $\n{u}$ are tangent. From this condition, one may also write the symmetry vector field in the following form,
\begin{equation}
    \n{u}=\frac{1}{|\n{B}|^2}(C\n{B}-\n{B}\times\nabla\Phi). \label{QSBu}
\end{equation}
Here $C=\n{B}\cdot\n{u}$ is some arbitrary function of space, as the constraint equation only describes the perpendicular part of $\n{u}$. \par
Because of magnetostatic equilibrium, we hav 
\begin{equation}
    \n{j}\times\n{B}=\nabla p \label{QSBp}
\end{equation}
where $\n{j}=\nabla\times\n{B}$ is the plasma current density. Because $\n{B}\cdot\nabla p=0$, it follows that $p=p(\psi)$, again assuming that $\n{B}$ is irrational. By continuity, we argue that this must be true for the rational cases as well (provided the rotational transform is not constant $\iota'\neq0$). \par
Because the plasma pressure is also a flux function, the current density $\n{j}$ also lives on magnetic flux surfaces. Furthermore, requiring $\nabla p$ to be nowhere vanishing (except the axis), the magnetic field and the currents are guaranteed to be non-collinear. Therefore, at every point on a flux surface, we may define a tangent plane spanned by $\n{j}$ and $\n{B}$. \par
Because $\n{u}$ also exists in this subspace, we can express it in the chosen basis as,
\begin{equation*}
    \n{u}=a\n{j}+\tilde{\beta}\n{B}
\end{equation*}
Taking the cross product of this form of $\n{u}$ with $\n{B}$, and making use of (\ref{QSBC2}) and (\ref{QSBp}), it is clear that $a=a(\psi)=-\Phi'/p'$, where the prime denotes a derivative with respect to the flux $\psi$. \par
Let us now apply (\ref{eqn:d.u}), namely $\nabla\cdot\n{u}=0$, on our form of $\n{u}$,
\begin{equation*}
    \nabla\cdot\n{u}=\n{B}\cdot\nabla\tilde{\beta}=0 ~~\therefore~~ \tilde{\beta}=\tilde{\beta}(\psi)
\end{equation*}
so that,
\begin{equation}
    \n{j}=-\frac{p'(\psi)}{\Phi'}\n{u}+\beta(\psi)\n{B}. \label{jub}
\end{equation}
Using the form (\ref{QSBu}), we can then rewrite the expression for $\n{j}$ in the form,
\begin{equation}
    \n{j}=\left(\beta(\psi)-\frac{p'(\psi)C}{B^2\Phi'}\right)\n{B}+\frac{p'(\psi)}{B^2}\n{B}\times\nabla\psi \label{jbetaC}
\end{equation}
The gauge-independent combination $C/\Phi'=(\mathbf{B}\cdot\nabla\psi\times\nabla B)/(\mathbf{B}\cdot\nabla B)$ here is the same as appears in (\ref{eqn:prePaulForm}), and it is a flux function. There we found a closed form for this combination in terms of the covariant representation of $\mathbf{B}$ in GBC. Equation (\ref{jbetaC}), together with $\nabla\cdot\mathbf{B}=0$ and $\mathbf{B}\cdot\nabla\psi=0$ constitute the whole set of governing equations for the QS, MHS field. The problem in this case can be formulated by providing $\beta$, $C/\Phi'$ and $p$ as input flux functions and solving for $\mathbf{B}$.  \par
Assuming that we have strong QS, $\beta=-C'/\Phi'$, obtaining an expression equivalent to that obtained in (\ref{eq:JB_JQ_def}). The assumption of strong QS was, however, not necessary when working with the coordinate representation. The adoption of some form of covariant representation of $\mathbf{B}$ seems to be necessary to obtain such a relation. To that end, GBC is convenient, though not necessary. In fact, [\onlinecite{burby2021}] employs the $B^\flat$ form (the generalisation of the covariant representation of the magnetic field), without specifying coordinates, to prove (\ref{eq:JB_JQ_def}). Note, however, that the difference in the form of $\beta$ is merely a matter of how the flux function is defined. The problem has $\beta$ or $\Phi'$ as free flux function inputs.

\bibliography{GBC}

\end{document}